\documentclass[11pt]{article}
\usepackage{epsf}

\newcommand{\bmat}{\left(\begin{array}}
\newcommand{\emat}{\end{array}\right)}
\def\NPB{Nucl. Phys. B}

\def\yzero{\smash{\hbox{$y\kern-4pt\raise1pt\hbox{${}^\circ$}$}}}

\def\b{\beta}

\def\beq{\begin{equation}}
\def\eeq{\end{equation}}
\def\beqa{\begin{eqnarray}}
\def\eeqa{\end{eqnarray}}

\def\-{\hphantom{-}}
\def\ov{\overline}
\def\s2{\frac{1}{\sqrt2}}

\def\beq{\begin{equation}}
\def\eeq{\end{equation}}
\def\beqa{\begin{eqnarray}}
\def\eeqa{\end{eqnarray}}

\def\IF{\relax{\rm I\kern-.18em F}}
\def\II{\relax{\rm I\kern-.18em I}}
\def\IP{\relax{\rm I\kern-.18em P}}
\def\IC{\relax\hbox{\kern.25em$\inbar\kern-.3em{\rm C}$}}
\def\IR{\relax{\rm I\kern-.18em R}}

\def\cp{{\cal P}}

\def\Dsl{\,\raise.15ex\hbox{/}\mkern-13.5mu D} 
\def\IZ{Z\kern-.4em  Z}

 \def\cp#1{\relax\ifmmode {\IP\kern-2pt{}_{#1}}\else $\IP\kern-2pt{}_{#1}$\=fi}

\catcode`\@=11
\newdimen\@rotdimen
\newbox\@rotbox

\def\@vspec#1{\special{ps:#1}}
\def\@rotstart#1{\@vspec{gsave currentpoint currentpoint translate
   #1 neg exch neg exch translate}}
\def\@rotfinish{\@vspec{currentpoint grestore moveto}}
\def\@rotr#1{\@rotdimen=\ht#1\advance\@rotdimen by\dp#1%
   \hbox to\@rotdimen{\hskip\ht#1\vbox to\wd#1{\@rotstart{90 rotate}%
   \box#1\vss}\hss}\@rotfinish}
\def\@rotl#1{\@rotdimen=\ht#1\advance\@rotdimen by\dp#1%
   \hbox to\@rotdimen{\vbox to\wd#1{\vskip\wd#1\@rotstart{270 rotate}%
   \box#1\vss}\hss}\@rotfinish}%
\def\@rotu#1{\@rotdimen=\ht#1\advance\@rotdimen by\dp#1%
   \hbox to\wd#1{\hskip\wd#1\vbox to\@rotdimen{\vskip\@rotdimen
   \@rotstart{-1 dup scale}\box#1\vss}\hss}\@rotfinish}%
\def\@rotf#1{\hbox to\wd#1{\hskip\wd#1\@rotstart{-1 1 scale}%
   \box#1\hss}\@rotfinish}%
\def\rotate{\@ifnextchar[{\@rotate}{\@rotate[l]}}
\def\@rotate[#1]#2{\setbox\@rotbox=\hbox{#2}\@nameuse{@rot#1}\@rotbox}

\catcode`\@=12

\begin{document}

\makeatletter \@addtoreset{equation}{section} \makeatother
\renewcommand{\theequation}{\thesection.\arabic{equation}}
\pagestyle{empty}
\pagestyle{empty}
\vspace{0.5in}
\rightline{FTUAM-02/30}
\rightline{IFT-UAM/CSIC-02-49}
\rightline{\today}
\vspace{2.0cm}
\setcounter{footnote}{0}

\begin{center}
\LARGE{
{\bf Standard Model Compactifications from Intersecting Branes}
\footnote{Contribution to the SUSY '02 proceedings}
}
\\[4mm]
{\large{ Christos ~Kokorelis 
\footnote{ Christos.Kokorelis@uam.es}
}
\\[1mm]}
\normalsize{\em Departamento de F\'\i sica Te\'orica C-XI and 
Instituto de F\'\i sica 
Te\'orica C-XVI,}
,\\[-0.3em]
{\em Universidad Aut\'onoma de Madrid, Cantoblanco, 28049, Madrid, Spain}
\end{center}
\vspace{1.0mm}


\begin{center}
{\small  ABSTRACT}
\end{center}

We discuss the construction
of four dimensional non-supersymmetric models obtained from configurations 
of D6-branes intersecting at angles.
We present the first examples of string GUT models
which break exactly
to the Standard Model (SM) at low energy. 
Even though the models are non supersymmetric (SUSY), the demand
that some open string sectors preserve N=1 SUSY creates gauge singlet 
scalars that break the extra anomaly free U(1)'s generically 
present in the models, 
predicting $s{\tilde \nu}_R$'s
and necessarily creating Majorana mass terms for
right handed neutrinos.

\newpage
\setcounter{page}{1} \pagestyle{plain}
\renewcommand{\thefootnote}{\arabic{footnote}}
\setcounter{footnote}{0}

\section{Introduction}
Recently, compactifications of intersecting D6-branes on a an 
orientifolded
D6-torus \cite{tessera}, which make use of the presence of a 
discrete antisymmetric B-field \cite{eksi1} to achieve
three generation models,
have received much attention. In this framework, 
it was possible to achieve the first unique examples
of
classes of four dimensional models with only the SM at low energy
 \cite{louis2, kokos1, kokos2, kokos3, kokos4}.
 Thus in \cite{louis2} the first systematic examples
 of models, based on a SM-like structure (SLM) at the
 string scale, with just the SM at low energy were
 constructed. Extended constructions
 of \cite{louis2} based on five and six stacks of
 SLM's, constructed as
 unique
 deformations around the quark intersection
 number structure of \cite{louis2}, appeared
 in \cite{kokos3, kokos4} respectively.\newline
 In this talk, we will review the construction of the
 first
 examples of string theory GUT models
 that break only to
 the SM at low energies \cite{kokos1}. The role of the
 extra branes,
 needed to satisfy the RR tadpole
 cancellation conditions, as well a construction
 of five stack GUT models may be found in
 \cite{kokos2}.

Also a partial list of other works in the context of intersecting
 branes can be seen in \cite{pente, franco,
 uran, ena1, ena}.

\section{Only the SM at low energy from GUTS of
intersecting D6-branes}
The GUT models that we will describe have some
important phenomenological properties e.g the proton is
stable, as the corresponding gauge boson
becomes massive and the baryon number survives as a
global
symmetry to low energies. Also the parameters of the
models \footnote{also parametrizing the solutions
to the RR tadpole cancellation conditions}
can easily accommodate small neutrino
masses of order of 0.1-10 eV in consistency with neutrino
oscillation experiments \cite{kokos1}.

The GUT construction that we will focus our
attention is based on the four stack
Pati-Salam like structure 
$U(4)_c \times U(2)_L \times U(2)_R
\times U(1)_d$ at the string scale or
$SU(4)_c \times SU(2)_L \times SU(2)_R \times U(1)_a 
\times U(1)_b \times U(1)_c \times U(1)_d$.
In intersecting brane worlds fermions get
localized in the intersections between branes.
The open string spectrum of the models can be seen in
table (\ref{spectrum8}).
\begin{table}[htb] \footnotesize
\renewcommand{\arraystretch}{1.5}
\begin{center}
\begin{tabular}{|c|c||c|c||c||c|c|}
\hline
Fields &Intersection  & $\bullet$ $SU(4)_C \times SU(2)_L \times SU(2)_R$
 $\bullet$&
$Q_a$ & $Q_b$ & $Q_c$ & $Q_d$ \\
\hline
 $F_L$& $I_{ab^{\ast}}=3$ &
$3 \times (4,  2, 1)$ & $1$ & $1$ & $0$ &$0$ \\
 ${\bar F}_R$  &$I_{a c}=-3 $ & $3 \times ({\ov 4}, 1, 2)$ &
$-1$ & $0$ & $1$ & $0$\\
 $\chi_L$& $I_{bd} = -12$ &  $12 \times (1, {\ov 2}, 1)$ &
$0$ & $-1$ & $0$ & $1$ \\    
 $\chi_R$& $I_{cd^{\ast}} = -12$ &  $12 \times (1, 1, {\ov 2})$ &
$0$ & $0$ & $-1$ &$-1$ \\\hline
 $\omega_L$& $I_{aa^{\ast}}$ &  $12 \b^2 {\tilde \epsilon} \times (6, 1, 1)$ &
$2{\tilde \epsilon}$ & $0$ & $0$ &$0$ \\
 $y_R$& $I_{aa^{\ast}}$ &  $6  \b^2 {\tilde \epsilon} \times ({\bar 6}, 1, 1)$ &
$-2{\tilde \epsilon}$ & $0$ & $0$ &$0$ \\
 $z_R$& $I_{aa^{\ast}}$ & $6  \b^2 {\tilde \epsilon}  \times ({\bar 10}, 1, 1)$ &
$-2{\tilde \epsilon}$ & $0$ & $0$ &$0$ \\
 $s_L$& $I_{dd^{\ast}}$ &  $24 \b^2 {\tilde \epsilon} \times (1, 1, 1)$ &
$0$ & $0$ & $0$ &$2{\tilde \epsilon}$ \\
\hline
\end{tabular}
\end{center}
\caption{\small Fermionic spectrum of the $SU(4)_C \times
SU(2)_L \times SU(2)_R$, type I models together with $U(1)$ charges.
The spectrum appearing in the full table is of
PS-A models of \cite{kokos1}. Note that the
representation context is considered by assuming
${\tilde \epsilon} = 1$. In the general case
${\tilde \epsilon} = \pm 1$.
If ${\tilde \epsilon} = -1$ then the conjugate
fields should be considered, e.g.
if ${\tilde \epsilon} = -1$, the
$\omega_L$ field should transform as $({\bar 6}, 1, 1)_{(-2, 0, 0, 0)}$.
\label{spectrum8}}
\end{table}
As can be seen in the bottom part of table
(\ref{spectrum8}) there are a number of exotic fermions
present in
the spectrum in the string scale. They will all
receive masses of the order of the string scale apart
from the $\chi_L$ fermions, which receive a mass of
order \footnote{where $\upsilon$ the scale of
electroweak symmetry breaking}
$\upsilon^2/M_s$.
For the models corresponding to the spectrum of table
(\ref{spectrum8}) the associated solutions to the RR
tadpoles may are seen in the first four rows of
table (\ref{ma101}).
The satisfaction of RR tadpole cancellation conditions
formulated as
\beqa
\sum_a n_a^1 n_a^2 n_a^3 = 16,\ \ \sum_a m_a^1 m_a^2 n_a^3 = 0,\ \
\sum_a m_a^1 n_a^2 m_a^3 = 0,\
\ \sum_a n_a^1 m_a^2 m_a^3 = 0
\eeqa
requires the presence of extra U(1) branes not
originally
present in the models, necessary to satisfy the RR
tadpoles \cite{kokos2}. The presence of extra branes
provides us with a mechanism
for generating gauge singlet scalars that may be used
to break the extra U(1)'s.
\begin{table}[htb]\footnotesize
\renewcommand{\arraystretch}{2}
\begin{center}
\begin{tabular}{||c||c|c|c||}
\hline
\hline
$N_i$ & $(n_i^1, m_i^1)$ & $(n_i^2, m_i^2)$ & $(n_i^3, m_i^3)$\\
\hline\hline
 $N_a=4$ & $(0, \epsilon)$  &
$(n_a^2, 3 \epsilon \b_2)$ & $({\tilde \epsilon}, {\tilde \epsilon}/2)$  \\
\hline
$N_b=2$  & $(-1, \epsilon m_b^1 )$ & $(1/\beta_2, 0)$ &
$({\tilde \epsilon}, {\tilde \epsilon}/2)$ \\
\hline
$N_c=2$ & $(1, \epsilon m_c^1 )$ &   $(1/\beta_2, 0)$  & 
$({\tilde \epsilon}, -{\tilde \epsilon}/2)$ \\    
\hline
$N_d=1$ & $(0, \epsilon)$ &  $(n_d^2, 6 \epsilon \b_2)$
  & $(-2{\tilde \epsilon}, {\tilde \epsilon})$  \\   
\hline
$\vdots$ & $\vdots$ & $\vdots$  & $\vdots$  \\
\hline
$N_{h}$ & $(1/\beta_1, 0)$ &  $(1/\beta_2, 0)$
  & $(2{\tilde \epsilon}, 0)$  \\\hline
\end{tabular}
\end{center}
\caption{\small
Tadpole solutions for PS-A type models with
D6-branes wrapping numbers giving rise to the 
fermionic spectrum and the SM,
$SU(3)_C \times SU(2)_L \times U(1)_Y$, gauge group at low energies.
The wrappings 
depend on two integer parameters, 
$n_a^2$, $n_d^2$, the NS-background $\beta_i$ and the 
phase parameters $\epsilon = {\tilde \epsilon }= \pm 1$. 
Also there is an additional dependence on the two wrapping
numbers, integer of half integer,
$m_b^1$, $m_c^1$. Note the presence of the
$N_{h}$ {\em extra} $U(1)$ branes.
\label{ma101}}
\end{table}
We note that the presence of extra branes alone is not
enough to make the fermions of table (1) massive.
The missing ingredient is the presence of N=1 SUSY in some
open string sectors. N=1 SUSY is not originally present in
the models. However, if we demand that certain sectors
preserve N=1 SYSY the tadpole parameters of table (2)
have enough freedom to
accommodate such a choice. Also, in the lack of
N=1 SUSY there is neither a Majorana coupling for the
right handed neutrinos, $\nu_R$'s, nor mass
terms for the fermions
of table (1).\newline
Now if we demand that the $ac$-sector preserves
N=1 SUSY, we pull out from the massive modes
the superpartner of the ${\bar F}_R$ antiparticles, e.g.
${\bar F}_R^B$ and thus a Majorana mass term for
$\nu_R$'s appears, e.g.
$F_R F_R {\bar F}_R^H {\bar F}_R^H$.
Also by demanding that the $dd^{\star}$ respects $N=1$
SUSY the gauge singlet scalar $s_L^H$ appears which may receive a
vev.
Also, a number of scalars are generically present in
the models including the left-right symmetry
breaking scalars
\beq
H_1 = (4, 1, 2)_{(1, 0, 1, 0)},\  \ H_2 = ({\bar 4}, 1,
{\bar 2})_{(-1, 0, -1, 0)}
\label{ele1}
\eeq
as well the electroweak symmetry breaking bidoublet scalars
\beq
h_1 = (1, 2, 2)_{(0, 1, 1, 0)},\  \ h_2 = (1, {\bar 2},
{\bar 2})_{(0, -1, -1, 0)}
\label{ele2}
\eeq
Moreover the presence of N=1 SUSY implies the relation
$2 n_a^2 =\ n_d^2$ and also some relations between the complex moduli parameters on
the factorizable orientifolded $T^6$.
Finally all fermions of table (1)
receive a mass. For example the 6-plet fermions
$\omega_L$ receive a mass from the coupling
\beq
\sim        
\langle H_1 \rangle \langle F_R^H \rangle
\langle  H_1 \rangle \langle F_R^H \rangle \sim M_s
\eeq
Also we note that the Yukawa term 
\beqa
F_L {\bar F}_R h,&& h=\{h_1, h_2\},
\label{yukbre}
\eeqa
is responsible for the electroweak 
symmetry breaking. This term is responsible for 
giving Dirac masses to up quarks and
neutrinos.
In fact, we
get
\beq
\lambda_1 F_L {\bar F}_R h  \rightarrow (\lambda_1 \  \upsilon)
(u_i u_j^c + \nu_i N_j^c) + (\lambda_1 \  {\tilde \upsilon}) 
\cdot (d_i d_j^c + e_i e_j^c),  
\label{era2}
\eeq
giving \footnote{The couplings
$\lambda_1$, $\lambda_2$ depend on the worldsheet area
between the D6 branes that cross at these interaction
vertices.}
non-zero tree level masses
to the fields present. 
These mass relations may be retained at tree level only, since as the model
has a non-supersymmetric fermion spectrum, it
breaks supersymmetry on the brane,
it will receive higher order
corrections.  
It is interesting that from (\ref{era2}) we derive the
well known GUT relation
\beq
m_d =\ m_e \ .
\label{gutscale}
\eeq
as well the ``unnatural''
 \beq
m_u =\ m_{N^c \nu} \ .
\label{gutscale1}
\eeq 
The latter is modified due to the presence of the Majorana
term for $\nu_R$'s leading us to a see-saw mechanism of the
Frogatt-Nielsen type \cite{kokos1}.
As a closing statement, we note that from the four
extra U(1)'s present at $M_s$, three become massive
in the presence
of a generalized Green-Schwarz mechanism involving
couplings of the U(1)'s to 
RR fields, while the
fourth U(1) could be broken either from $s_L^B$ or from
extra
scalars involved in the presence of N=1 SUSY
in sectors coming from the mixing between
the U(1) $d$-brane and the extra $N_h$
branes \cite{kokos2}.

\small{}

\end{document}